\documentclass[fleqn,usenatbib]{mnras}

\makeatletter
\def\@abstract{%
  \list{}{%
    \listparindent\realparindent
    \itemindent\z@
    \labelwidth\z@
    \labelsep\z@
    \leftmargin11pc
    \rightmargin\z@
    \parsep 0pt plus 1pt}%
  \item[]%
  \reset@font\normalsize{\bf ABSTRACT}\\\reset@font\abslarge
}
\def\@keywords{%
  \list{}{%
    \labelwidth\z@
    \labelsep\z@
    \leftmargin11pc
    \rightmargin\z@
    \parsep 0pt plus 1pt}%
  \item[]\reset@font\abslarge{\bf Key words: }%
}
\makeatother

\usepackage{newtxtext,newtxmath}

\usepackage[T1]{fontenc}

\usepackage{amsmath}
\usepackage{graphicx}

\usepackage{hyperref}
\hypersetup{colorlinks=true, citecolor=blue, linkcolor=blue}

\def\aj{Astron.\ J. }
\def\apj{Astrophys.\ J. } 
\def\apjl{Astrophys.\ J. }
\def\apjs{Astrophys.\ J.\ (Supp.) }
\def\aap{Astron.\ Astrophys. }

\def\prd{Phys.\ Rev.\ D }

\newcommand{\orcid}[1]{\protect\href{https://orcid.org/#1}{\protect\includegraphics[width=8pt]{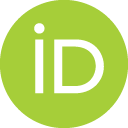}}}

\def\vv#1{\boldsymbol{#1}}

\def\H{\overline H}
\def\K{J}
\def\alp{\alpha}
\def\cte{\mathrm{const}}
\def\vr{\vv{r}}
\def\vp{\vv{p}}
\def\vL{\vv{L}}
\def\vS{\vv{S}}
\def\vK{\vv{\K}}
\def\ur{\vv{\hat r}}
\def\vk{\vv{k}}
\def\vs{\vv{s}}
\def\ui{u}
\def\vi{v}
\def\xi{x_i}
\def\x{x_1}
\def\y{x_2}
\def\z{z}
\def\dphi{\Delta \phi}
\def\zx{X(\x,\ui)}
\def\zxs{X^2(\x, \ui)}
\def\gx{S(\x,\ui)}
\def\uic{\ui_c}
\def\deltai{\theta_{12}}

\def\bfx#1{#1}

\def\figpath{}
\def \llabel#1{\label{#1}}

\begin{document}


\title{Cassini states for black hole binaries}

\author[A. C. M. Correia]{
Alexandre C. M. Correia$^{\orcid{0000-0002-8946-8579}1,2}$\\
$^{1}$ CIDMA, Departamento de F\'{i}sica, Universidade de Aveiro, Campus de Santiago, 3810-193 Aveiro, Portugal\\
$^{2}$ ASD, IMCCE-CNRS UMR8028, Observatoire de Paris, 77 Av. Denfert-Rochereau, 75014 Paris, France
}

\date{Accepted 2015 December 8. Received 2015 November 5; in original form 2015 September 25}

\maketitle

\begin{abstract}
Cassini states correspond to the equilibria of the spin axis of a body when its orbit is perturbed.
They were initially described for planetary satellites, but the spin axes of black hole binaries also present this kind of equilibria.
In previous works, Cassini states were reported as spin-orbit resonances, but actually the spin of black-hole binaries is in circulation and there is no resonant motion.
Here we provide a general description of the spin dynamics of black hole binary systems based on a Hamiltonian formalism.
In absence of dissipation the problem is integrable and it is easy to identify all possible trajectories for the spin for a given value of the total angular momentum.
As the system collapses due to radiation reaction, the Cassini states are shifted to different positions, which modifies the dynamics around them.
This is why the final spin distribution may differ from the initial one.
Our method provides a simple way of predicting the distribution of the spin of black hole binaries  at the end of the inspiral phase. 
\end{abstract}


\section{Introduction}

Following observations of the Moon, 
\citet{Cassini_1693} established three empirical laws on its rotational motion.
The first stated that the rotation rate and the orbital mean motion are synchronous, the second that the angle between Moon's equator and the ecliptic is constant, and the third that the Moon's spin axis and the normals to its orbital plane and ecliptic remain coplanar.
The observed physical librations are described as departures of the rotational motion from these three equilibrium laws. 
\citet{Colombo_1966} has shown that the second and third laws are independent of the first one, 
and generalised these laws to any satellite or planet whose nodal line on the invariant plane shifts because of perturbations. 
In his approach, the Hamiltonian of a slightly aspherical body is developed in a reference frame that precesses with the orbit.
If the angular momentum and the energy are approximately conserved, the precession of the spin axis relative to the coordinate system fixed in the orbital plane is determined by the intersection of a sphere and a parabolic cylinder.
The spin axis is fixed relative to the precessing orbit when the energy has an extreme value.
Thus, these equilibria states for the spin axis can be the end point of dissipation, 
and they received the name of Cassini states \citep{Colombo_1966, Peale_1969, Ward_1975, Correia_2015}.

\citet{Schnittman_2004} has found that spinning black hole binaries can also present stable configurations where the two spin axes and the orbital angular momentum vector remain coplanar. 
Because of the loss of energy and orbital angular momentum through gravitational radiation reaction, the spins may end in libration around these equilibria. 
The final spin evolution of black holes is particularly interesting, since the spin alignment during the inspiral phase can change significantly the distribution of black hole recoil velocities \citep{Kesden_etal_2010a, Kesden_etal_2010b, Bogdanovic_etal_2007, Gupta_Gopakumar_2014, Gerosa_etal_2013, Gerosa_etal_2015a, Berti_etal_2012}.
One of the most difficult aspects of studying the spinning black hole binary system is the problem of visualizing and analyzing the orientation of the two spins and the angular momentum in an informative way.
Previous studies always describe the libration around the coplanar configurations as spin-orbit resonances  \citep{Schnittman_2004, Kesden_etal_2010a, Berti_etal_2012, Gerosa_etal_2013, Gerosa_etal_2014, Kesden_etal_2015, Gerosa_etal_2015b, Gerosa_etal_2015a}.
\bfx{
However, this description is not consistent with the expectation that the two normal modes of this problem become resonant, since for black hole binaries one frequency is usually smaller than the other \citep{Racine_2008}.
Indeed, by a suitable variable transformation, one can change the critical argument from libration to circulation. 
More generally, one can say there is resonant motion only if there is a clear change in the topology of the phase space, with a separatrix between the circulation and libration regions \citep[see][]{Henrard_Lemaitre_1983}.}

In this Letter we revisit the spin dynamics of black hole binaries adopting an Hamiltonian formalism.
In absence of dissipation, the problem is integrable and we provide a simple analytical method to find its solutions.
We show
that the equilibria found by \citet{Schnittman_2004} is similar to the one observed by \citet{Cassini_1693} for the Moon.

\section{Secular dynamics}

We consider a binary composed of two black holes with relative position $\vr$, masses $m_1$ and $m_2$, and spins $\vS_1$ and $\vS_2$, respectively. 
The inspiral of the system is governed by radiation reaction at binary separations $ r = ||\vr|| < 10^4$~$r_g$, where $r_g = GM/c^2$ is the gravitational radius.
Numerical-relativity simulations with initial separations $r/r_g > 10 $ are still too computationally expensive.  
However, in the range $10 < r/r_g < 10^4$, 
the binary dynamics can be described using the spinning Taylor-expanded PN Hamiltonian. 
As in previous studies \citep{Schnittman_2004, Kesden_etal_2010a, Berti_etal_2012, Gerosa_etal_2013, Gerosa_etal_2014, Kesden_etal_2015, Gerosa_etal_2015b, Gerosa_etal_2015a}, we focus our analysis in this range.
In the barycenter frame, the Hamiltonian depends on the canonical variables ($\vr, \vp$) and on the spins vectors.
For the purposes of our analysis, it is sufficient to restrict the discussion to the Newtonian contribution, $H_N$, and include only the leading 1.5PN spin-orbit interaction, $H_{SO}$, and the leading 2PN interaction, $H_{SS}$, which includes spin-induced monopole-quadrupole terms \citep{Barker_OConnell_1975, Damour_2001, Buonanno_etal_2011}.
The full Hamiltonian then reads $H = H_N + H_{SO} + H_{SS}$, where
\begin{equation}
H_N = \frac{\vp^2}{2 \mu} - \frac{G M \mu}{r} \ ,
\llabel{150821a}
\end{equation}
\begin{equation}
H_{SO} = \frac{2 G}{c^2 r^3} \vS_e \cdot \vL \ ,
\llabel{150821b}
\end{equation}
\begin{equation}
H_{SS} = \frac{G \mu}{2 M c^2 r^3} \left[ 3 \left(\vS_0 \cdot \ur \right)^2 - \vS_0^2 \right] \ ,
\llabel{150821c}
\end{equation}
with $\mu = m_1 m_2 / M$, $M= m_1 + m_2$, $q=m_2/m_1$, $\ur = \vr / r$, $\vL = \vr \times \vp$, 
\begin{equation}
\vS_0 = \left(1+ q\right) \vS_1 + \left(1+ 1/q \right) \vS_2 \ ,
\llabel{150821d}
\end{equation}
\begin{equation}
\vS_e = \left(1+ \frac{3q}{4} \right) \vS_1 + \left(1+ \frac{3}{4q} \right) \vS_2 \ .
\llabel{150821e}
\end{equation}

The evolution in the PN limit occurs on three distinct time scales: the orbital time $t_{o} \sim (r^3/G M)^{1/2}$, the spin precession time $t_{p} \sim t_{o} (M/\mu) (r / r_g) $, and the radiation-reaction time on which the orbital angular momentum decreases $t_{rr} \sim t_{p} (r/r_g)^{3/2}$.
Therefore, for $r/r_g > 10$ we have $t_{o} \ll t_{p} \ll t_{rr}$, which means that 1) we can neglect the effect of radiation-reaction over one precession cycle; 2) we can average the Hamiltonian over one orbital period to obtain the secular motion of the spin \citep{Correia_etal_2011}: 
\begin{equation}
\H = -\frac{G M \mu}{2 a} +
4 \alp \, \vS_e \cdot \vL 
-\frac{\alp \mu}{2 M} \left[ \frac{3}{L^2}\left(\vS_0 \cdot \vL \right)^2 - \vS_0^2 \right] 
\ , \llabel{150821f}
\end{equation}
where $\alp = G/(2 c^2 r_0^3)$, $r_0 = a \sqrt{1-e^2}$, $L = ||\vL|| = \mu \sqrt{G M a (1-e^2)}$, $a$ is the semi-major axis, and $e$ is the eccentricity.
In the secular conservative problem, all quantities appearing in the Hamiltonian (\ref{150821f}) are constant, except for the angular momentum components.
The evolution of the system can therefore be obtained from the Hamiltonian through Poisson brackets \citep{Dullin_2004, Tremaine_etal_2009}
\begin{equation}
\dot \vS_i  = \{ \vS_i, \H \} = \frac{\partial \H}{\partial \vS_i} \times \vS_i \ ,
\quad
\dot \vL = \{ \vL, \H \} = \frac{\partial \H}{\partial \vL} \times \vL 
\ , \llabel{150821g}
\end{equation}
which gives
\begin{equation}
\dot \vS_i = \alp \left[ \left(4 + \frac{3 m_j}{m_i} - \frac{3 \mu}{m_i} \lambda \right) \vL + \vS_j \right] \times \vS_i
\ , \llabel{150821i}
\end{equation}
\begin{equation}
\dot \vL = \alp \left[ 4 \, \vS_e - \frac{3 \mu}{M} \lambda \vS_0 \right] \times \vL
\ , \llabel{150821h}
\end{equation}
with $i \ne j = 1, 2 $, and $\lambda = \vS_0 \cdot \vL / L^2 $.
From previous expressions we also see that in absence of radiation reaction 1) the norms of the individual angular momentum vectors are conserved; 2) the total angular momentum vector is also conserved:
\begin{equation}
\vK = \vL + \vS_1 + \vS_2 = \cte \ .
\llabel{141216i}
\end{equation}

\section{Reduced problem}

The equations of motion can be simplified if we consider only the relative position in space of the unit vectors $\vs_i = \vS_i/S_i$ and $\vk = \vL/ L $, given by the direction cosines \citep{Schnittman_2004, Goldreich_1966a, Boue_Laskar_2006} 
\begin{equation}
\xi = \cos \theta_i = \vs_i \cdot \vk \ , \quad 
z = \cos \deltai = \vs_1 \cdot \vs_2 \ ,
\llabel{141216j}
\end{equation}
together with the ``berlingot'' shaped volume
\begin{equation}
w = \vk \cdot (\vs_1 \times \vs_2) = \pm \sqrt{1- \x^2 - \y^2 - z^2 + 2 \x \y z} 
\ . \llabel{141216k}
\end{equation}
The equations of motion (\ref{150821i}) and (\ref{150821h}) become
\begin{equation}
\dot \xi = (-1)^i \beta S_j \frac{\mu \lambda - M}{m_j} \ , \quad
\dot z = \beta L \left(\frac{\Delta m}{M} \lambda - \frac{\Delta m}{\mu} \right) \ ,
\llabel{141216l}
\end{equation}
with $\beta = 3 \alp \, w $, $\Delta m = m_1 - m_2$, and
\begin{equation}
\dot w =  (\y z-\x) \frac{\dot \x}{w} + (\x z-\y) \frac{\dot \y}{w} + (\x \y -z) \frac{\dot z}{w} \ .
\llabel{141216m}
\end{equation}
We can get $w$ directly from expression (\ref{141216k}), although this last equation can be useful for determining whether $w$ is positive or negative.
In addition, we still have two remaining integrals\footnote{For equal masses ($\Delta m = 0$) the angle $\deltai$ between the spin axes is also conserved (Eq.\,(\ref{141216l})).}, one from the total angular momentum (\ref{141216i})
\begin{equation}
\K_0 = L \, S_1 \x + L \, S_2 \y + S_1 S_2 z = \cte
\ , \llabel{141216o}
\end{equation}
and another from the Hamiltonian (\ref{150821f}). 
Thus, equations (\ref{141216l}) reduce to an integrable problem \citep{Boue_Laskar_2006}.
From expression (\ref{141216o}) we can write
\begin{eqnarray}
\vS_0^2 &=& 2 (2+q+1/q) S_1 S_2 z + (1+q)^2 S_1^2 + (1+1/q)^2 S_2^2  \nonumber \\
&=& - 2 (2+q+1/q) L (S_1 \x + S_2 \y) + \cte
\ . \llabel{151104a}
\end{eqnarray}   
Replacing in expression (\ref{150821f})  gives for the Hamiltonian 
\begin{equation}
\H_0 =
3 \alp L^2 \lambda -\frac{3 \alp L^2 \mu}{2 M} \lambda^2 = \cte
\ , \llabel{150828a}
\end{equation}
hence we conclude that 
\begin{equation}
\lambda (\x,\y) = \frac{M}{L} \left( \frac{S_1}{m_1} \, \x +  \frac{S_2}{m_2} \, \y \right) 
\ , \llabel{150828b}
\end{equation}
is also a conserved quantity \citep[see also][]{Racine_2008}.

\section{Precessional motion}

\llabel{newvars}

\begin{figure*}
\includegraphics[width=\textwidth]{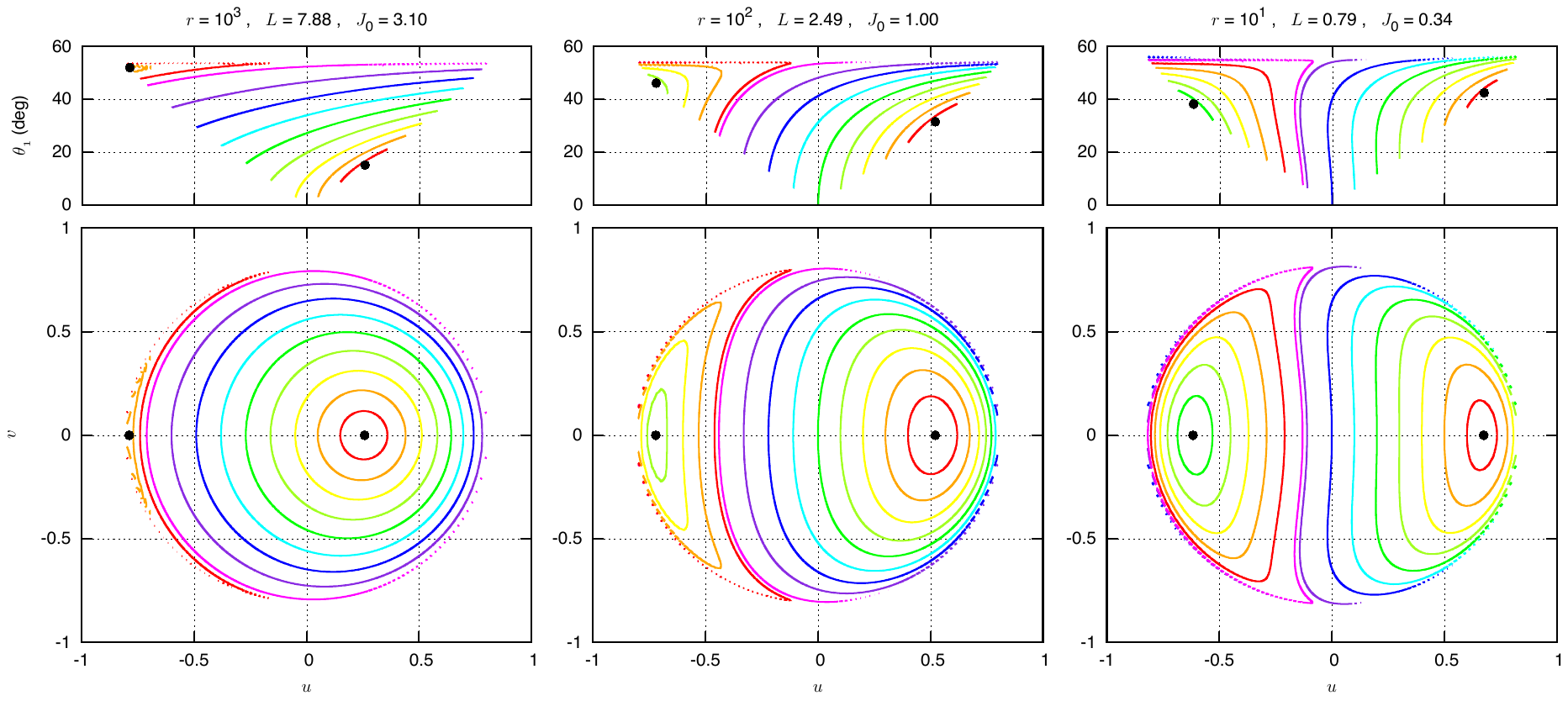} 
 \caption{Secular trajectories for the spin of the more massive object in a black hole binary with $q=0.9$ and $\chi_1 = \chi_2 = 1$ (the units are $G=M=c=1$).
Each panel shows the spin evolution at different binary separations, $r$, for which all trajectories have the same total angular momentum $\K_0$ (Eq.\,(\ref{141216o})), they only differ by the value of $\lambda$ (Eq.\,(\ref{150828b})).
We show the spin projected on the orbit normal (top), and its projection on the orbital plane (bottom). Cassini states are marked with a dot. 
 \llabel{contours}  }
\end{figure*}

The spin evolution is better described by the projection of the spin axis in the orbital plane ($\ui,\vi$), obtained as
\begin{equation}
\ui = \frac{\z - \x \y}{\sqrt{1-\y^2}} = \sin \theta_1 \cos \dphi \ , 
\llabel{141217a}
\end{equation}
and
\begin{equation}
\vi = \frac{w}{\sqrt{1-\y^2}} = \sin \theta_1 \sin \dphi \ ,
\llabel{141217b}
\end{equation}
where $\dphi$ is the angle measured along the orbital plane from the projection of $\vs_1$ to the projection of $\vs_2$ \citep[see Fig.~1 in][]{Gerosa_etal_2015a}. 
Thus, when $\sin \dphi = 0$ the unit vectors ($\vs_1$, $\vs_2$, $\vk$) lie in the same plane.
With this choice, $\x$ only depends on the new variables\footnote{We considered that $\x>0$, but this method is still valid for $\x<0$ adopting $\x=-\sqrt{1 - \ui^2 - \vi^2}$. It also stands for $\y$ by switching the indexes $_1$ and $_2$. However, if $x_1$ and $x_2$ simultaneously oscillate around 0, the validity of this method is not assured.}
\begin{equation}
\x = \sqrt{1 - u^2 - v^2}  \ . \llabel{141217c}
\end{equation}
We can get $\z$ from spherical trigonometry
\begin{equation}
\z = \x \y + \ui \sqrt{1 - \y^2} \ , \llabel{141217d}
\end{equation}
while $\y$ can be obtained by eliminating $\z$ in expression (\ref{141216o}) using the previous identity:
\begin{equation}
(L + S_1 \x) \, \y + S_1 \, \ui \sqrt{1 - \y^2} = (\K_0 - L \, S_1 \x)/S_2
\ , \llabel{141217e}
\end{equation}
which can be explicitly solved for $\y$ as
\begin{equation}
\y = \frac{(L + S_1 \, \x) \zx - S_1 \, \ui \sqrt{1-\zxs}}{\gx}
\ , \llabel{141217f}
\end{equation}
with
\begin{equation}
\zx = \frac{\K_0 - L \, S_1 \x}{S_2 \, \gx} 
\ , 
\llabel{141224a0}
\end{equation}
\begin{equation}
\gx = \sqrt{(L + S_1 \, \x)^2+(S_1 \, \ui)^2}
\ . \llabel{141224a}
\end{equation}
Therefore, $\y$ depends only on ($\x, \ui$), hence on the new variables ($\ui,\vi$), as well as $\lambda$ (Eq.\,(\ref{150828b}))
\begin{equation}
\lambda (\x, \y) = \lambda (\x, \ui,\K_0) = \lambda (\ui,\vi,\K_0) \ . \llabel{141217g}
\end{equation}

In Figure~\ref{contours} we show the secular trajectories for the spin of the more massive object in a black hole binary with $q=0.9$ at different binary separations (we adopt $S_i = \chi_i m_i^2$, with $\chi_i=1$).
In each panel, all trajectories have the same total angular momentum, 
obtained with initial $\x=z=1$ and $\y=0.5$ (Eq.\,(\ref{141216o})).
They only differ by the value of $\lambda$, that corresponds to different initial orientations of the spin vectors (Eq.\,(\ref{150828b})).
The level curves, obtained directly from expression (\ref{141217g}) without integrating the equations of motion, fully characterise the spin dynamics for a given separation and total angular momentum.

We observe that the spin is always in circulation around a fix point (Cassini state).
Previous studies \citep{Schnittman_2004, Kesden_etal_2010a, Gerosa_etal_2013, Gerosa_etal_2015a} report that when the angle $\dphi$ librates there is resonant motion.
However, we clearly see there is no resonance in this problem, since there is no separatrix emerging from a fix point.
Because the fix points are displaced from the origin ($\ui = 0$), for trajectories where we always have $u>0$ (or $u<0$), it appears that $\dphi$ librates around $0$ (or $\pi$).


\section{Cassini states}

Cassini states correspond to equilibria of the spin axis.
Thus, they are given by the extrema of the Hamiltonian (\ref{150828a}):
\begin{equation}
\frac{\partial \lambda}{\partial \ui} = 0 \quad \wedge \quad \frac{\partial \lambda}{\partial \vi} = 0 \ .
\llabel{150513a}
\end{equation} 
Since $\lambda = \lambda (\x, \ui)$, we have
\begin{equation}
\frac{\partial  \lambda}{\partial \vi} = \frac{\partial  \lambda}{\partial \x}\frac{\partial \x}{\partial \vi} = - \frac{\partial  \lambda}{\partial \x}\frac{\vi}{\x} = 0 \ .
\llabel{150514a}
\end{equation} 
We then conclude that $\vi = 0$ is always a possible equilibrium solution (equivalent to $\dphi = 0$ or $\pi$), where the unit vectors $\vs_1$, $\vs_2$, and $\vk$ remain coplanar.
Using $\vi=0$ in (\ref{150513a}) gives an implicit condition for the coplanar states, $\uic= \sin \theta_c $:
\begin{equation}
\tan \theta_c =  \frac{\uic}{x_c} =  \frac{m_1}{m_2} \, \frac{S_2}{S_1} \left. \frac{\partial \y}{\partial \uic} \right|_{\vi=0} 
\ , \llabel{150511a}
\end{equation}
where the derivative is computed using expression (\ref{141217f}) with $\ui=\uic$ and $\x = x_c = \sqrt{1-\uic^2}$.
The roots of (\ref{150511a}) can be found in the interval $ \uic \in [-1,1]$ using numerical methods or simply by plotting its graph.

Alternatively, coplanar states can be obtained as stationary solutions for the equations of motion for which $\vi = 0$ \citep{Schnittman_2004}.
Therefore, they can also be obtained by setting $w = 0$ and $\dot w = 0$ (Eq.\,(\ref{141216m})), that is\begin{equation}
\uic = \frac{(\mu \lambda - M) \left[ (x_c z-\y) \frac{S_1}{m_1} - (\y z- x_c)  \frac{S_2}{m_2} \right]}{L \left(\frac{\Delta m}{M} \lambda - \frac{\Delta m}{\mu} \right) \sqrt{1 - \y^2} }  \ ,
\llabel{141217h}
\end{equation}
where $\z$ and $\y$ are obtained from (\ref{141217d}) and  (\ref{141217f}) with $\ui=\uic$ and $\x = x_c$.

In Figure~\ref{cassinistates} we plot the Cassini states as a function of the binary separation for the same system shown in Figure~\ref{contours}.
We observe that there are always two equilibrium points $\uic = (\ui_0, \, \ui_\pi)$, one corresponding to an aligned configuration for the spins ($\ui_0>0$ and $\dphi=0$), and another corresponding to an anti-aligned configuration ($\ui_\pi<0$ and $\dphi=\pi$).
For $r \gg 10^4 $, we have $\ui_0 \rightarrow 0$, hence the angle $\dphi$ appears to always circulate around zero.
As the binary separation decreases, both $\uic$ increase.
Therefore, for trajectories initially circulating around $\ui_0$ (or $\ui_\pi$) the average obliquity of the spin increases (or decreases). 
Moreover, for the trajectories around $\ui_0$, the angle $\dphi$ appears to switch from circulation to libration as $\ui_0$ moves away from the origin (see Fig.\,\ref{secevol}).

\begin{figure}
\begin{center}
\includegraphics[width=\columnwidth]{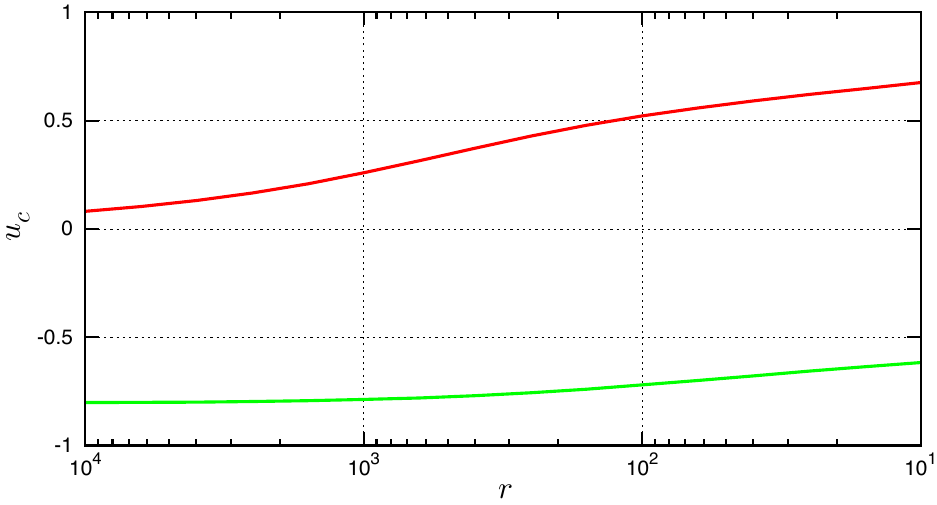} 
 \caption{Cassini states as a function of the binary separation $r$. 
These equilibria are obtained by solving equation (\ref{150511a}). 
The vertical lines correspond to the configurations shown in Figure~\ref{contours}. \llabel{cassinistates}  }
\end{center}
\end{figure}

The secular spin-orbit problem for black hole binaries has two independent frequencies (normal modes), $\nu_0$ and $\nu_\pi$.
In the case of similar masses ($q \approx 1$), we have \citep{Racine_2008}
\begin{equation}
\nu_0 \approx  \alp ||\vK|| \Big( 7 - \frac32 \lambda \Big)
 \ , \quad 
\frac{\nu_\pi}{\nu_0} \approx  1 - \frac{12 - 3 \lambda}{14 - 3 \lambda} \frac{||\vS||}{||\vK||} 
\ , \llabel{150915a} 
\end{equation}
where $\nu_0$ can be seen as the precession rate of the total spin $\vS = \vS_1 + \vS_2$ about the total angular momentum $\vK$ (or the precession of $\vL$ about $\vK$), and $\nu_\pi$ as a nutation frequency.

Any projection of the spin $S_p(t)$ can thus be written as
\begin{equation}
S_p (t) = \sum_{j,k} A_{jk} \, \mathrm{e}^{i (j \nu_0 + k \nu_\pi) t}
\approx A_{00} + A_{10} \, \mathrm{e}^{i \nu_0 t} + A_{01} \, \mathrm{e}^{i \nu_\pi t}
\ , \llabel{150916a}
\end{equation}
where the $A_{jk}$ are determined by the initial conditions.
We have $|A_{10}| \ga |A_{01}|$ for trajectories circulating around the Cassini state $\ui_0$ 
(or $|A_{10}| \la |A_{01}|$ for those around $\ui_\pi$),
with $|A_{01}|=0$ at the equilibrium point $\ui_0$ (or $|A_{10}|=0$ at  $\ui_\pi$).
Spin-orbit resonances can occur whenever $\nu_0 = \nu_\pi$.
\bfx{
For black hole binaries resonant motion is unlikely since we usually have $\nu_\pi < \nu_0$.
For lower mass ratios ($q \ll 1$), the dynamics is dominated by state $\ui_0$ and there is also no resonant motion.
However, we cannot rule out any extreme example where the system would actually exhibit formal resonant behavior.
In that case, the analytical model discussed in previous sections is still valid.
We would see appearing a separatrix and additional Cassini states,} as it happens in the classic planetary case \citep[see][]{Correia_2015}.

\section{Long-term evolution}

In the previous analysis we neglected the radiation reaction damping of the system through gravitational wave emission.
Of course, in all physically realistic black hole binaries, gravitational radiation plays a major role in the secular evolution of the orbit. 
With the inclusion of radiation reaction, the orbital angular momentum and the eccentricity evolve according to \citep[e.g.][]{Peters_1964, Gergely_etal_1998}
\begin{equation}
\dot \vL = - \frac{32}{5} \frac{G^{7/2} M^{5/2} \mu^2}{c^5 a^{7/2}(1-e^2)^2} \left( 1 + \frac78 e^2 \right) \vk
\ , \llabel{150914a}
\end{equation}
\begin{equation}
\frac{\dot e}{e} = - \frac{304}{15} \frac{G^3 M^2 \mu}{c^5 a^4 (1-e^2)^{5/2}} \left( 1 + \frac{121}{304} e^2 \right)
\ . \llabel{150914b}
\end{equation}
The full problem is no longer integrable, so we need to perform numerical simulations of the previous equations together with (\ref{150821i}) and (\ref{150821h}) to track the spin evolution of the system.
In Figure~\ref{secevol} we plot the evolution of the spin of the same system shown in Figure~\ref{contours} starting with initial $a = 10^3$, $e=0$, $\vi=0$, and $\ui=-0.73$ (left) or $\ui=-0.05$ (right).
For initial $\ui=-0.73$, the spin is initially dominated by the Cassini state $\ui_0$. 
Around $r \approx 600$, it switches to state $\ui_\pi$, since for $q=0.9$ this state dominates the majority of the trajectories with $\ui < 0$ at small separations (Fig.~\ref{contours}).
For initial $\ui=-0.05$, the spin is always dominated by the Cassini state $\ui_0$.
The equilibrium value for this state increases for small $r$ (Fig.~\ref{cassinistates}), so the trajectories \bfx{moving around it} also increase its average obliquity.
As a result, the angle $\dphi$ appears to switch from circulation to libration at $r \approx 500$ because $\ui$ becomes always positive for smaller separations.

\begin{figure*}
\begin{center}
\includegraphics[width=\textwidth]{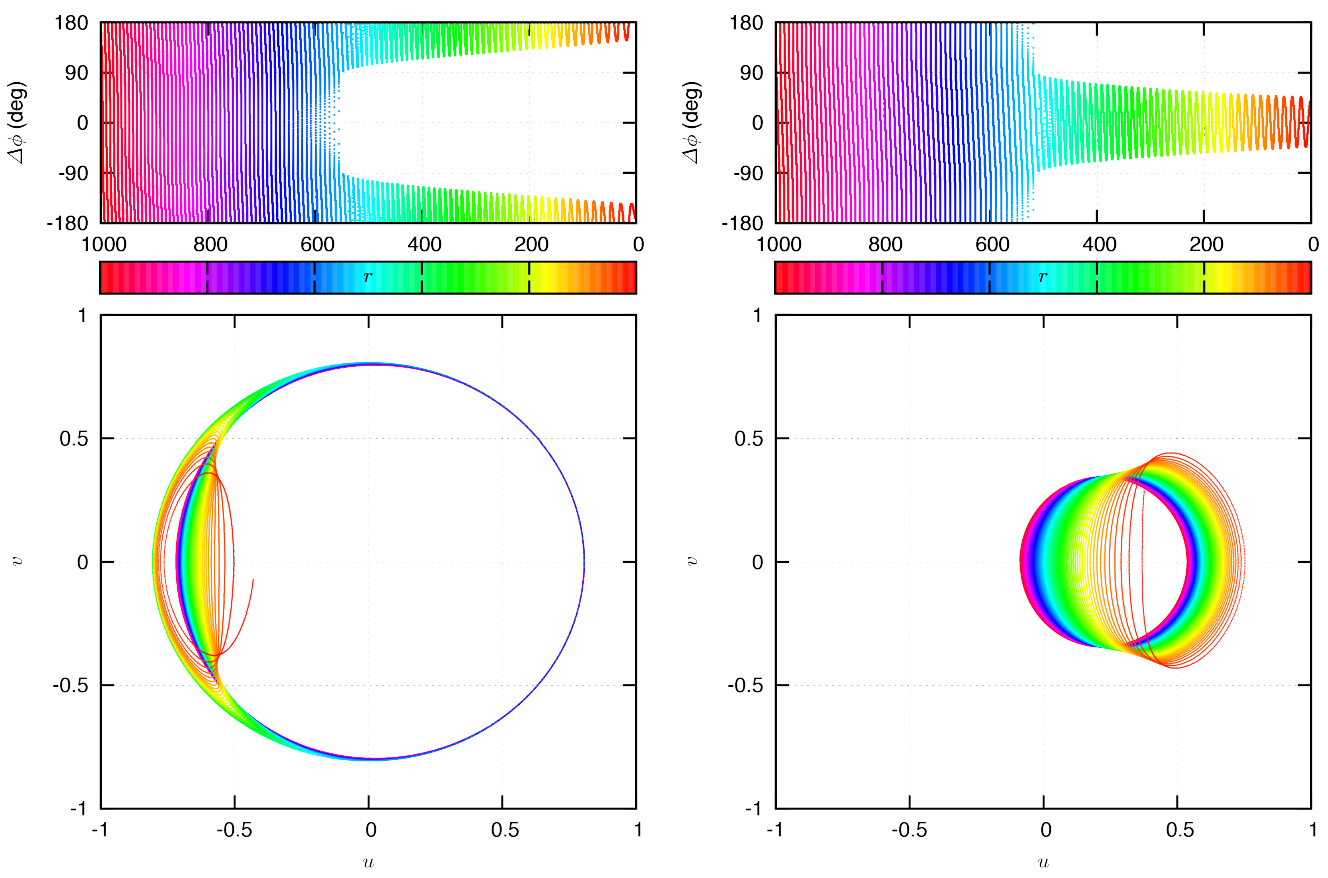} 
 \caption{Long-term spin evolution of the more massive object in a black hole binary with $q=0.9$ and $\chi_1 = \chi_2 = 1$. 
The initial conditions are $a = 10^3$, $e=0$, $\vi=0$, and $\ui=-0.73$ (left) or $\ui=-0.05$ (right).
We show the angle $\dphi$ (top) and the spin projected on the orbital plane (bottom).
The color scale is related to the binary separation. \llabel{secevol}  }
\end{center}
\end{figure*}

\section{Discussion}

In this Letter we presented a simple method for determining the equilibrium states and the secular trajectories for the spins of black hole binaries with separations $r/r_g > 10$.
For a given value of the total angular momentum, we are able to plot the global spin dynamics, including the equilibria points (Cassini states).
Our method only depends on the geometry of the Hamiltonian and thus does not require an integration of the equations of motion.
In addition, it allows us to correctly understand the dynamical regimes and to exclude the occurrence of spin-orbit resonances in this problem.
For black hole binaries, all trajectories actually circulate around the Cassini equilibria.
As the binary separation decreases due to radiation reaction, the spin dynamics is also modified.
However, since dissipation only occurs in the orbital angular momentum, the changes in the spin are a result of the modification in the phase space of the system.
Therefore, by looking at the initial position of the spin in the phase space and how the phase space is modified for small separations, it becomes possible to predict the final spin geometries at the end of the inspiral phase. 
This work is important to understand the distribution of black hole recoil velocities.

\subsection*{Acknowledgements}
We acknowledge support from CIDMA strategic project UID/MAT/04106/2013.

\bibliographystyle{mnras}
\bibliography{correia}

\end{document}